\documentclass[iop,apj]{emulateapj}
\usepackage{amsmath,amssymb,amstext}

\usepackage{color} 

\usepackage{natbib}
\slugcomment{as of \today}

\shorttitle{BSS in 47 Tuc}
\shortauthors{Parada et al.}

\usepackage{float} 
\DeclareGraphicsExtensions{,.pdf,.jpg}
\usepackage{subfigure}

\usepackage{hyperref}

\usepackage{colortbl}

\makeatletter

    \def\CT@@do@color{%
      \global\let\CT@do@color\relax
            \@tempdima\wd\z@
            \advance\@tempdima\@tempdimb
            \advance\@tempdima\@tempdimc
    \advance\@tempdimb\tabcolsep
    \advance\@tempdimc\tabcolsep
    \advance\@tempdima2\tabcolsep
            \kern-\@tempdimb
            \leaders\vrule
                    \hskip\@tempdima\@plus  1fill
            \kern-\@tempdimc
            \hskip-\wd\z@ \@plus -1fill }
    \makeatother

\usepackage{longtable}
\usepackage{rotating}
\usepackage{multirow}

\usepackage{array}
\newcolumntype{x}[1]{%
>{\centering\hspace{0pt}}p{#1}}
\newcolumntype{y}[1]{%
>{\raggedright\hspace{0pt}}p{#1}}

\usepackage[table]{xcolor}
\definecolor{light-gray}{gray}{0.85}
\begin{document}

\title{Formation and Evolution of Blue Stragglers in 47 Tucanae}

\author{Javiera Parada\altaffilmark{1}, Harvey Richer\altaffilmark{1}, Jeremy Heyl\altaffilmark{1},Jason Kalirai\altaffilmark{2}, Ryan Goldsbury\altaffilmark{1}}
\altaffiltext{1}{Department of Physics \& Astronomy, University of
  British Columbia, Vancouver, BC, Canada V6T 1Z1;
  jparada@phas.ubc.ca, heyl@phas.ubc.ca, richer@astro.ubc.ca  }
\altaffiltext{2}{Space Telescope Science Institute,Baltimore MD
  21218; Center for Astrophysical Sciences, Johns Hopkins University, Baltimore MD, 21218; jkalirai@stsci.edu }

\begin{abstract}
Blue stragglers (BSS) are stars whose position in the Color-Magnitude Diagram (CMD) places them above the main sequence turn-off (TO) point of a star cluster. Using data from the core of 47 Tuc in the ultraviolet (UV), we have identified various stellar populations in the CMD, and used their radial distributions to study the evolution and origin of BSS, and obtain a dynamical estimate of the mass of BSS systems. 
When we separate the BSS into two samples by their magnitude, we find that the bright BSS show a much more centrally concentrated radial distribution and thus higher mass estimate (over twice the TO mass for these BSS systems), suggesting an origin involving triple or multiple stellar systems. 
In contrast, the faint BSS are less concentrated, with a radial distribution similar to the main sequence (MS) binaries, pointing to the MS binaries as the likely progenitors of these BSS. 
Putting our data together with available photometric data in the visible and using MESA evolutionary models, we calculate the expected number of stars in each evolutionary stage for the normal evolution of stars and the number of stars coming from the evolution of BSS. The results indicate that BSS have a post-MS evolution comparable to that of a normal star of the same mass and a MS BSS lifetime of about 200-300 Myr. We also find that the excess population of asymptotic giant branch (AGB) stars in 47 Tuc is due to evolved BSS.\\

\end{abstract}

\maketitle


\section{INTRODUCTION} \label{sec:intro}

With the development of high resolution astronomical imaging, astronomers have been able to study globular star clusters (GC) in great detail, exposing the presence of different anomalous stellar populations. An important example of such stars are blue stragglers (BSS). First discovered by \cite{sandage} in the GC M3, BSS were described as an extension of the main sequence (MS) defying normal stellar evolution within a cluster. How these stars are formed in GC and where they go after they leave their MS stage has been a constant debate (\cite{eco_bss}, especially Chapter 9 and 11). 

One of the largest populations of BSS resides in the GC NGC 104 (47 Tucanae, 47 Tuc). Although 47 Tuc has been the target of many investigations, observations using ultraviolet (UV) filters, such as the one used to obtain the current data set, facilitate the selection of BSS and make them one of the brightest populations in the CMD.


In the last two decades BSS have been found in many GCs \citep{ferraro2012} as well as open clusters \citep{deMarchi2006,ahumada}, in dwarf galaxies \citep{santana} and in the field of our galaxy \citep{santucci}. In older stellar clusters, there is generally no evidence of recent star formation episodes, thus, for these stars to look brighter and bluer than the turnoff they had to go through some rejuvenating process.The BSS formation mechanisms can be divided in many different ways but they all must comply with two main conditions: {\it i)} there must be at least one MS star involved, and {\it ii)} one of the stars involved must gain mass in order to become rejuvenated. In fact, the positions of BSS on the CMD suggest that these stars are in fact more massive than the TO stars. The first attempt to directly measure the mass of a BSS was done by \cite{shara}, studying one of the brightest BSS in the core of 47 Tuc. They found a mass of $1.7 \pm 0.4 M_{\odot}$, almost twice the cluster TO mass of $\sim 0.9M_{\odot}$ \citep{hesser,thompson}. Later, different studies, including some done on variable BSS, have yielded masses between $\sim 1$ and $2 M_{\odot}$ for BSS in different GCs \citep{gilliland,demarco}. Recent results for pulsating BSS have provided a lower upper limit of $\sim 1.5M_{\odot}$ \citep{fiorentino2014,fiorentino2015}. 


There are two possible ways for a star to gain mass: mass transfer or merger. We will separate the initial scenarios into three different categories following the divisions chosen by \cite{bookCh11}:  {\it i)} direct collisions of stars \citep{hills}, {\it ii)} stellar evolution of primordial binaries \citep{McCrea}, and {\it iii)} dynamical evolution of hierarchical triple systems \citep{iben}. 

The last scenario became more important with the discovery of triple systems harbouring BSS (see \cite{s1082} for example), and the disagreement between the observed BSS populations and that obtained from combined N-body and stellar evolution simulations that considered only collisions and primordial binary evolution. \cite{perets2009} claimed that previous BSS formation studies demanded a fraction of the primordial binaries to be short period binaries. A previous publication by \cite{shortperiod} had shown that such systems actually come from longer period binaries that have been perturbed by a third star via the Lidov-Kozai mechanism \citep{lidov,kozai}. In fact, studies done on short period \citep{tokovinin} and contact \citep{pribulla} binaries showed that at least 40\% of these systems have distant companions.  

Recent studies following the formation channel proposed by \citeauthor{perets2009}, indicate that the Lidov-Kozai mechanism has a 21\% efficiency when it comes to forming  tight binaries \citep{naoz2014}. And, when applied to GC systems, it can contribute up to 10\% of the total BSS population \citep{antonini2015}. This population should show some observational differences when comparing them to the BSS with different origins. For instance their mass could reach much higher values than binary mass transfer scenarios, where part of the mass of the system is left in the WD companion. The WD is also another difference as BSS from a triple system are more likely to be left with a MS companion \citep{bookCh11}. 

Which mechanism dominates in the different environments in which BSS live is still under debate. Although no definite answer has been reached most studies agree that the observed populations today are a result of a combination of all the formation channels, with one mechanism prevailing over the others depending on the system's properties. Attempts to find the dominating formation mechanism in different GCs, have been based on finding the strongest correlation between the number of BSS and parameters of the cluster, like total or core mass, binary fraction and collision rate. \cite{knigge2009} found a strong correlation between the number of BSS in the core and the core mass in GCs. With the results pointing towards a binary origin for BSS, researchers started to look for confirmation of the correlation between BSS frequency and binary fraction already found by \cite{sollima2008} in low density GCs. \cite{milone_binaries} reaffirmed this correlation for a sample of 59 GCs. \cite{leigh2013} also tried to find a relation between binaries and BSS but their results showed a much stronger correlation with the core mass as found by \cite{knigge2009}, despite the fact that binary fraction in GCs anticorrelates with core mass \citep{milone2008}. One of the latest studies that included dynamical effects and stellar and binary evolution yielded {\it ``a dependence of blue straggler number on cluster mass, a tighter correlation with core mass, a weak dependence on the collisional parameter, and a strong dependence on the number of binary stars"} \citep{sills2013}. 

Observational evidence supporting the fact that more than one mechanism for BSS formation can take place has also been found. 
\cite{ferraroM30}, \cite{dalessandro2013} and \cite{simunovic2014} found two sequences of BSS in M30, NGC 362 and NGC 1261 respectively, in both clusters a blue and a red sequence of BSS was visible on the CMD. Both authors suggest that this feature is a possible consequence of two very distinct formation mechanisms taking place in the same cluster. In the case of M30, the blue-BSS sequence matches collisional models \citep{sills2009}, while the red-BSS sequence agrees with binary evolution models \citep{xin2015}.

\subsection{Blue Stragglers in 47 Tucanae} \label{subsec:intro_bss}

In the particular case of 47 Tuc, the study of its population of BSS started with the discovery of 21 such stars in one of the first HST observations of the core of this cluster \citep{paresce1991}. This small sample of BSS already provided evidence that the density of BSS is higher in the central regions of the cluster. Many investigations on the topic have taken place since then, \cite{sills2000} modelled the formation rate of BSS using data outside the core. The results obtained by these authors suggested that 47 Tuc may have stopped making BSS several billion years ago. The cluster underwent an epoch of enhanced BSS formation around the same time, and this was possibly connected to the epoch of primordial binary burning.

\cite{ferraro2004} discovered a bimodal radial distribution for the BSS in 47 Tuc, as seen in other GC like M3 \citep{ferraroM3}, and M55 \citep{lanzoni2007}. These distributions show a peak in the cluster center, decreasing at intermediate distances from the center, to rise again in the outskirts. \cite{mapelli2004} tried to reproduce the BSS radial distribution in 47 Tuc by choosing different formation mechanisms: collisional BSS in the innermost region and primordial binary evolution outside the core. The best representation of the observational data was obtained when 25\% of the BSS came from binaries and 75\% from collisions within $0.5r_{c}$ ($r_{c} = 1$ core radius). This result was later refined by \cite{mapelli2006} obtaining a best fit when 46\% of the BSS come from mass transfer and 54\% from collisions. The models were also able to predict the minimum in the radial distribution and its surrounding regions named by \cite{mapelli2004} as the {\it ``zone of avoidance"}, with the condition that external mass transfer BSS production began beyond $30r_{c}$.

Later on, \cite{monkman2006}, tried to explain the bimodal distribution with a purely collisional model throughout the cluster. Their results agreed with those found by \cite{mapelli2004,mapelli2006} for the core of the cluster where the collisional model represents the observational data. For their middle region (between 23 and 130 arcseconds from the center) BSS formation would have needed to stop about half a billion years ago. But for the external regions the collisional models were not able to predict the BSS population, a result that they concluded is likely due to another formation mechanism dominating the outskirts of 47 Tuc.

Around the same time the formation mechanisms debate was taking place, researchers found evidence that BSS in the core of 47 Tuc have masses larger than twice the MS TO mass. One result that suggested the presence of massive BSS was found by \cite{mclaughlin2006}, while studying the proper motion and dynamics of the cluster core. They determined that the velocity dispersion of BSS was smaller than that of the cluster giants by a factor of $\sqrt{2}$  (i.e. twice their mass). That same year, \cite{knigge2006} identified a detached binary system consisting of a $1.5M_{\odot}$ BSS primary with an active, upper MS companion. These massive BSS can only be the outcome of a process involving at least three progenitors.


Another interesting area of research is the evolution of BSS. In 1994 \cite{bailyn}, studied the central regions of 47 Tuc and found an overabundance of stars in the AGB of the cluster. He concluded that these extra stars could come from the evolution of BSS. \cite{beccari2006} also found evidence of an overabundance of stars in the AGB, which, according to theoretical tracks used by the authors, correspond to massive stars currently undergoing their RGB phase. Additionally \cite{beccari2006} separated the HB into faint and bright HB stars finding that both the overabundace of stars in the AGB and the presence of a bright extension of the HB could be related to the evolution of binary systems. Recently \cite{ferraro2016} confirmed observationally the presence of a massive star in the region of the CMD slightly brighter than the bulk of the HB. \cite{ferraro2016} report a mass of $1.4M_{\odot}$, much higher than the turnoff mass of the cluster, strongly suggesting this star is the result of the evolution of a BSS. 

\section{Observations and Analysis} \label{sec:obs_ana}

Using the Wide Field Camera 3 (WFC3) on board the Hubble Space Telescope (HST), data from the core of 47 Tuc were taken using two of the most ultraviolet filters, F225W and F336W whose central wavelengths are 235.9 nm and 335.9 nm respectively. The images covered up to a radial distance of $\sim 160$ arcseconds from the center of the cluster. A more detail description of the observations can be found in \cite{paper1} (hereafter {\bf paper1}). The data reduction and photometry were performed following the procedure described in \cite{kalirai2012}.

Additionally, we took photometric data available from the ACS (Advanced Camera for Surveys) Survey of Galactic Globular Cluster \citep{sarajedini2007}. For 47 Tuc, these data (hereafter the ACS data) cover only $\sim 105$ arcseconds radius from the center of the cluster, but uses filters in the visible range (F606W and F814W) that will help us to identify different evolutionary stages that do not have clear sequences in the UV.

The UV data (hereafter the WFC3 data), were corrected for incompleteness using artificial star tests in both the F225W and F336W images, obtaining a completeness rate as a function of magnitude and radial distance. This process is described in {\bf paper1} and explained in detail in \cite{heyl-diffusion}. We find that for stars above an F225W magnitude of 21 the completeness rate is close to unity. In the case of the ACS data, we do not correct for incompleteness as we are only using the bright stars in these filters and the completeness in this part of the CMD is also close to unity.

\section{Stellar Populations selection} \label{sec:selection}

The boundaries of the regions for the selection of each stellar population were chosen with the help of MESA (Modules for Experiments in Stellar Astrophysics; \citeauthor{paxton1} \citeyear{paxton1}, \citeyear{paxton2}, \citeyear{paxton3}) evolutionary models. The models were created using the pre-build \textsf{1M\_pre\_ms\_to\_wd} model in the test suite. The initial parameters were set to the appropriate values for 47 Tuc, with TO masses of $0.9M_{\odot}$ (corresponding to the cluster's TO mass), and additionally $1.1M_{\odot}$, $1.4M_{\odot}$ and $1.8M_{\odot}$, corresponding to masses for different BSS populations. Also, a metallicity of $Z = 4 \times 10^{-3}$ and helium abundance of $Y=0.256$ \citep{mesapar} are used. A detailed description of the construction of the models can be found in \cite{massloss}.

Slight modifications on the limits of those regions in the CMD show no effect in the resulting cumulative radial distributions and the number of stars within the regions does not change by more than a few percent.

\subsection{Main Sequence Binaries}
\label{subsec:sel_MSBn}

\begin{figure*}[ht]
	\centering
	\includegraphics[width=7.1in]{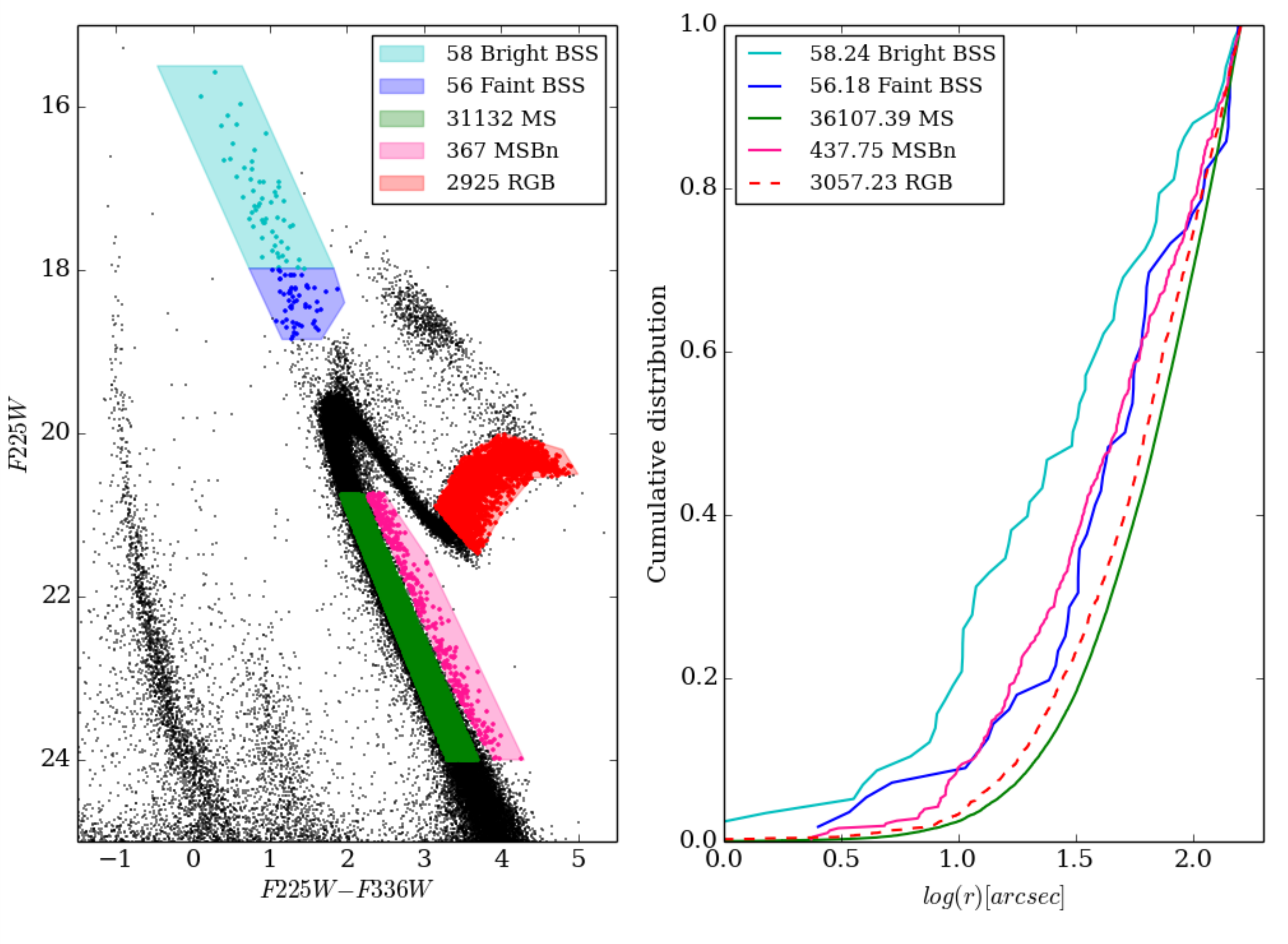}
	\caption{{\it Left:}$F225W,F225W-F336W$ CMD showing the selected stars for the faint and bright BSS, RGB, MS and MSBn populations. {\it Right:} Radial distributions for the selected samples. The legend on the CMD has the number of stars before correcting for incompleteness, while the legend on the right plot gives the size of the sample after correcting for incompleteness. The division between bright and faint BSS was chosen by maximizing the difference between their radial distributions in order to obtain two distinct BSS populations.}
	\label{fig:pop_sel}   
\end{figure*}

In an attempt to identify the population of stars responsible for the formation of BSS, we will later compare the BSS distribution against the binary star distribution. We have selected a sample of main sequence binaries (MSBn) that we expect to be mostly nearly equal mass binaries\footnote[1]{Equal mass binaries will be $\sim 0.75$ magnitudes brighter than their single MS counterpart, and will have the same colour as its components, placing them above the MS}. Both populations are shown in Figure \ref{fig:pop_sel}. The MSBn selection box starts at a faint magnitude of 24 and extends up to a magnitude a 20.7, with an almost constant width of 0.4 magnitudes (width reduces at the brighter end of this selection box to avoid contamination by SGB stars) containing a total of 367 stars. This number goes up to 438 after correcting for incompleteness.

We have also included a selection of stars on the main sequence (MS) to have a reference for the analysis of this population. To ensure there is limited contamination to the MSBn sample, a minimun distance of $0.2$ magnitudes is kept between the binary sequence and its single star sequence counterpart. Looking at the right panel of Figure \ref{fig:pop_sel}, we can see that the cumulative radial distribution for the MSBn is much more centrally concentrated than that of the single MS stars.

\subsection{Blue Stragglers}
\label{subsec:sel_BSS}
It can be seen in Figure ~\ref{fig:pop_sel} that the BSS population is easily identify on the UV CMD as an extension of the MS of the cluster. Starting a few tenths of magnitudes above the turn-off point and extending for almost 4 magnitudes, the total number of BSS in the sample contains almost 150 stars. For this study, we have decided to exclude the very faint BSS, and have taken only those that are at least $\sim 0.7$ magnitudes brighter than the TO, to avoid any possible contamination due to blends. This decision was also based on the fact that when we plot the BSS sample on the ACS CMD, the fainter BSS on the UV sample are very close to the $F606W,(F606W-F818W)$ TO, almost blending with the MS. It is important that we have clean BSS samples for both data sets. 

After delimiting the BSS sample we end up with 114 BSS, which we divide into two sub-samples, faint and bright BSS, each containing half of the total stars. When we divided the sample in half we noticed a big difference in the radial distributions with the bright BSS (bBSS) looking more centrally concentrated than the faint BSS (fBSS).

\subsection{Reference Population}
\label{subsec:sel_ref}
To trace the distribution of cluster stars we selected the RGB as the reference population. Although previous studies \citep{ferraro2003, ferraro2004} indicate the HB as the most natural reference population in the UV CMDs, due to this branch being well separated from other branches, we are concerned with the contamination of the HB by AGB stars and evolving BSS. We discuss this later in section \ref{sec:discussion}.

In the UV, specifically with the filters chosen for this work, the RGB is well defined and easy to identify in the CMD. Even though the RGB is not separated from the SGB, the shape of the CMD makes it easy to get a clean sample. It is important to have a clean sample of RGB stars as we will later use the number of stars in this region to estimate the expected number of stars in the various evolutionary stages. To make sure there is no contamination of SGB stars in our RGB sample, we start our RGB box a few tenths of a magnitude above and to the right of the end of the SGB.  Figure ~\ref{fig:pop_sel} (red) shows the final selection for the RGB with a total of 2925 stars before completeness corrections and $\sim3050$ stars after the corrections.

As mentioned in previous sections, we will compare our data to the ACS sample. On the ACS CMD, the RGB, especially in the fainter part of this branch, is also well defined. Cross matching the selection of the RGB on the UV CMD to the visible CMD, also gets us a clean RGB sample of stars starting around $\sim 0.5$ magnitudes brighter than the TO, which tells us that our efforts to exclude SGB stars from our UV sample were successful. Even at the bright end of the RGB on the visible CMD we can see that this branch is well separated from the horizontal and asymptotic giant branches, making it a suitable reference population also in these filters. Because the ACS field is smaller than the WFC3 field, we also expect our RGB sample to be smaller, coming to a total of $\sim 2200$ stars compared to the $\sim 3000$ we had before.

\subsection{ACS Data Selection}
\label{subsec:sel_ata}

\begin{figure}[H]
	\centering
	\includegraphics[width=5.7in]{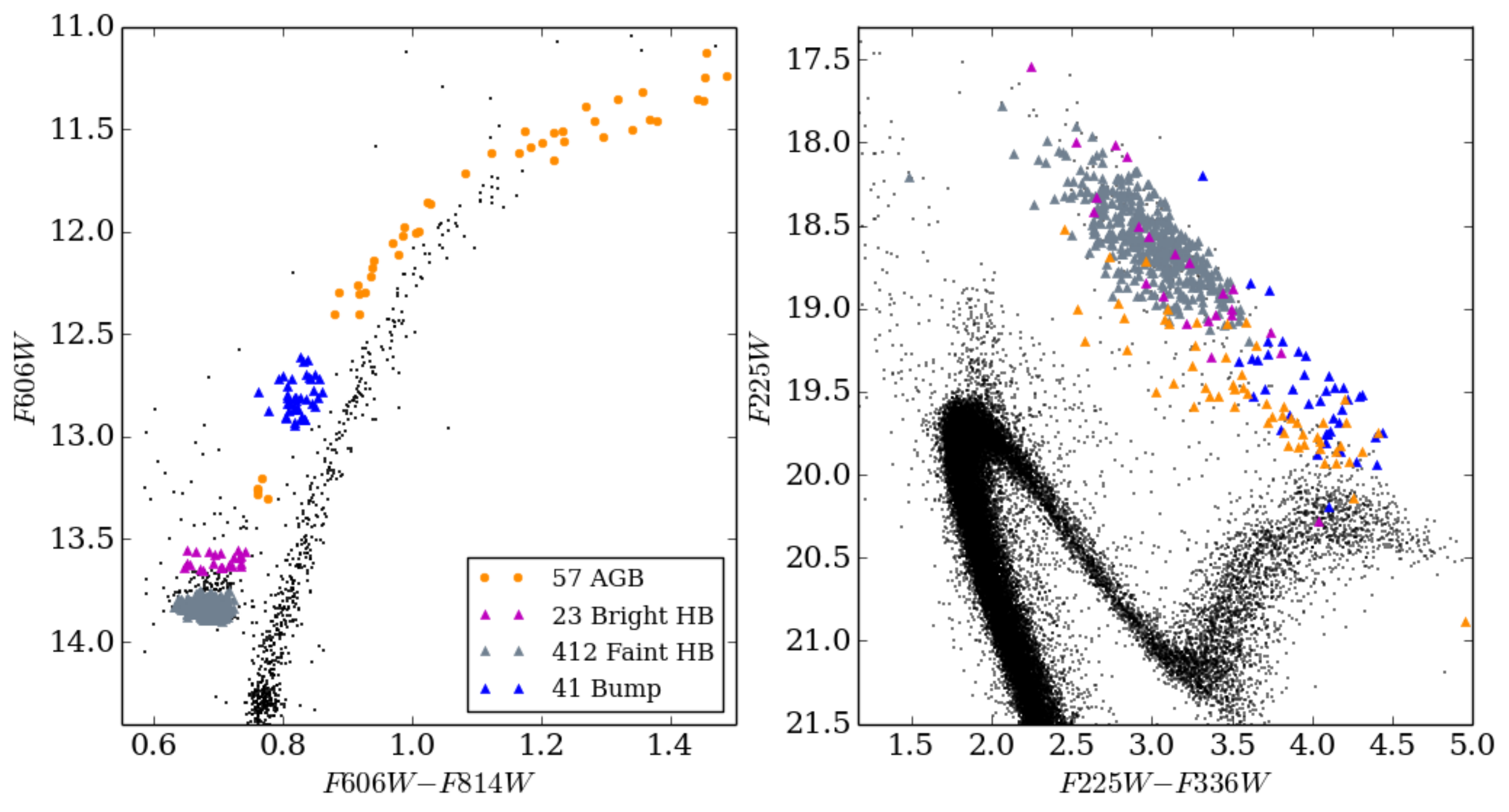}
	\caption[$F606W,F606W-F814W$ CMD showing the selection of the stellar populations on the ACS data, and where they fall on the $F225W,F225W-F336W$ CMD as evidence of contamination to the UV HB]{{\it Left:} $F606W,F606W-F814W$ CMD with the selection of the stellar populations on the ACS data. {\it Right:} $F225W,F225W-F336W$ CMD showing where the stars selected on the ACS data fall on the UV CMD. We can see a clear contamination of the UV HB by stars in stellar evolutionary stages different from the normal HB but that form clear branches on the ACS CMD. The number of stars in each sample is given in the inset in the left plot.}
	\label{fig:ata_ref}   
\end{figure}

As in the WFC3 CMD, we used MESA models to choose our regions which are shown in Figure \ref{fig:ata_ref}. Following a similar procedure as the one used by \cite{beccari2006} in this same cluster, we separate the HB into faint and bright HB stars leaving a small magnitude gap between them. Additionally we separate the AGB stars isolating the AGB bump from the rest of the stars in this evolutionary stage. The detailed reasoning behind the division of faint and bright HB stars and the difference between the AGB and the bump on this branch will be explained in sections \ref{sec:results} and \ref{sec:discussion}. The important point for now is confirming the presence of contaminating stars on the HB of the UV CMD pictured on the right panel of Figure \ref{fig:ata_ref}. Using the same color code on each panel we can see how the bright HB, AGB and the stars on the bump of the AGB picked on the ACS CMD fall in the same region as the HB stars on the WFC3 CMD. 

Although we lose some stars as we had to reduce our field size to match the ACS field, we can identify where the AGB from the evolution of normal stars falls in the UV CMD and obtain a cleaner sample of HB stars. This leads to the proper classification of over 100 stars that we would have otherwise needed to ignore.

In order to be able to compare the ACS and WFC3 data sets, the data for $F225W$ and $F336W$ were reduced to the same field as the one covered by the $F606W$ and $F814W$ which is also in the core but expands to a radius of only 105 arcseconds. When using the ACS field, every star included in the analysis had to be measured in all four filters.

\onecolumngrid

\twocolumngrid


\section{RESULTS} \label{sec:results}

\subsection{Blue Stragglers}
\label{subsec:res_BSS}
We have already pointed out the difference between the masses of faint and bright BSS. When looking at the cumulative radial distribution of the two regions of BSS in Figure \ref{fig:pop_sel} we can also see a significant difference between the two samples. According to the KS-test results, faint and bright BSS have only a 1.0\% probability of being drawn from the same population and are significantly different from the reference population with {\it p}-values of $0.03$ for fBSS and $\sim 10^{-6}$ for the bBSS.

From a visual examination of the radial distribution plot we noticed a similarity between the fBSS and the MSBn. This is confirmed by the KS-test which yields a {\it p}-value of $0.76$ suggesting a possible relation between these two groups of stars that is not present between the bBSS and MSBn ({\it p}-value = 0.01). The KS-test results between the regions highlighted in Figure \ref{fig:pop_sel} are summarized in table \ref{table:kstest-bss}.

\begin{table}[ht]\centering
 \caption{KS-test {\it p}-value results between the populations selected in Figure \ref{fig:pop_sel}. There is no apparent relation between the populations except between the MSBn and the fBSS.}
 \centering
 \begin{tabular}{ x{38pt}  x{38pt}  x{38pt}  x{38pt}  x{38pt}} 
& & & & \multicolumn{1}{c}{\cellcolor{light-gray}\bf bBSS}      	 \tabularnewline \cline{5-5}

& & & \multicolumn{1}{c}{\cellcolor{light-gray}\bf fBSS} & 0.01 \tabularnewline \cline{4-5}

& & \multicolumn{1}{c}{\cellcolor{light-gray}\bf MS} &  $\sim 10^{-4}$  & $\sim 10^{-8}$ \tabularnewline \cline{3-5}

& \multicolumn{1}{c}{\cellcolor{light-gray}\bf MSBn} & $\sim 10^{-13}$  & 0.76 &  0.01  \tabularnewline \cline{2-5}

\multicolumn{1}{c}{\cellcolor{light-gray}\bf RGB} & $\sim 10^{-5}$ & $\sim 10^{-17}$   &  $0.03$ & $\sim 10^{-6}$ \tabularnewline \cline{1-5}
 \end{tabular}
  \label{table:kstest-bss}
\end{table}

\subsection{Evolved Blue Stragglers}
\label{subsec:res_eBSS}

\begin{figure*}[ht]
	\centering
	\includegraphics[width=5.7in]{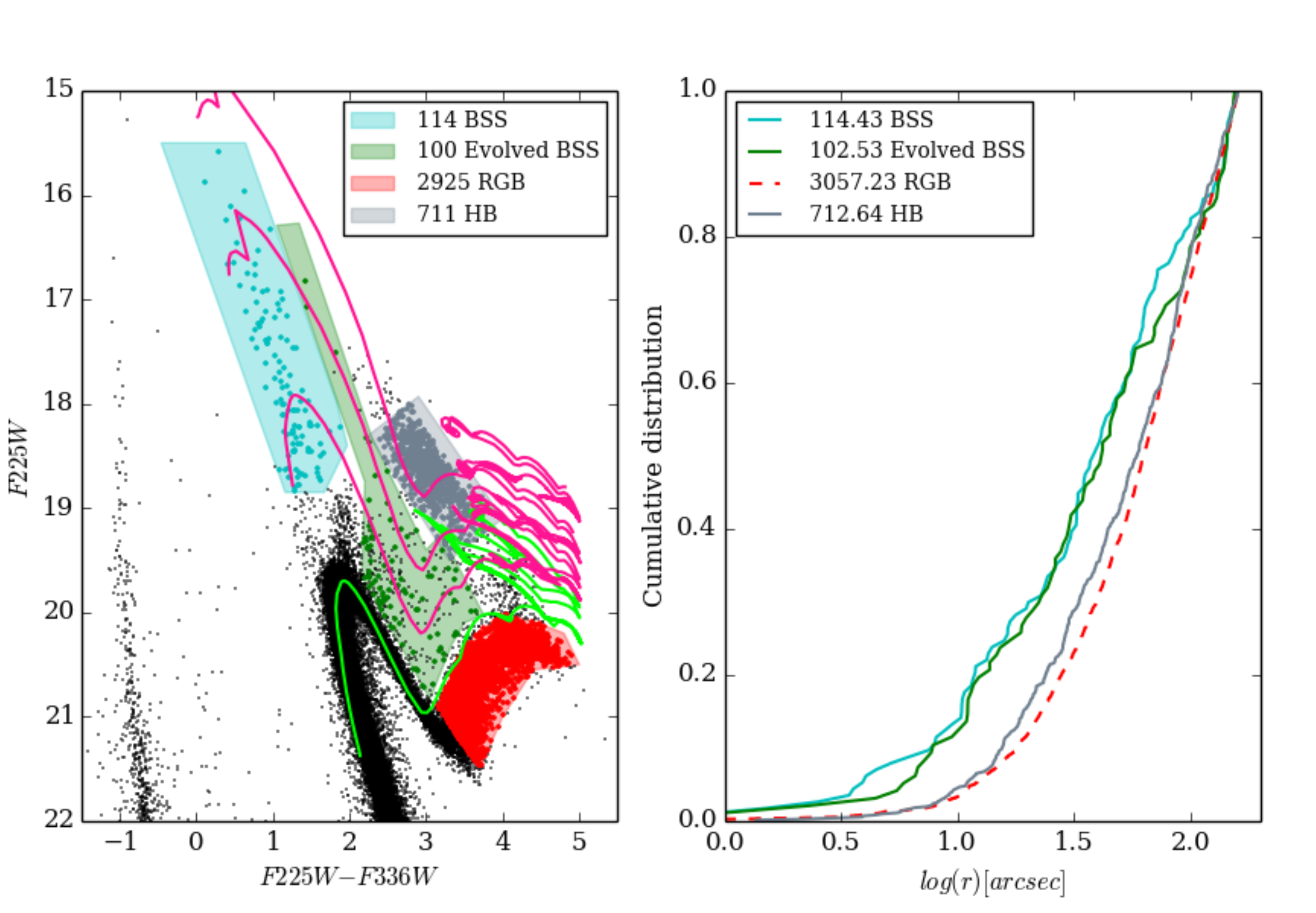}
	\caption{{\it Left:}UV CMD, the green and magenta curves are the MESA evolutionary models for stars with initial masses of $0.9M_\odot$, $1.1M_\odot$, $1.4M_\odot$, and $1.8M_\odot$ from bottom to top. {\it Right:} Radial distributions for the selected samples on the CMD. As on previous figures, the legend on the CMD has the number of stars before correcting for incompleteness, while the legend in the right plot gives the size of the sample after correcting for incompleteness.}
	\label{fig:eBSS}   
\end{figure*}

Distinguishing between the various evolutionary stages on the $F225W,F225W-F336W$ CMD beyond the RGB is complicated. Although the HB seems to be clear, the number of stars and the radial distribution of this branch disagrees with the models suggesting an over abundance of stars. In Figure \ref{fig:eBSS}, left panel, we can display the upper part of the CMD along with four MESA evolutionary models. The lowest mass model, $0.9M_\odot$, shows the evolution for a star with a mass approximately equal to the 47 Tuc TO mass. According to this model, the RGB lasts for $\sim4.2\times10^{8}$ years while the HB only $\sim0.7\times10^8$ years. 
Considering that all the stars going through the RGB phase come from the evolution of non-BSS stars, the number of RGB stars ($3060$) predicts $510$ HB stars and $110$ AGB stars. This gives us an excess of around 100 observed stars in the HB region, as specified in table \ref{table:times-wfc3}. 

Looking at table \ref{table:times-wfc3}, it can be seen that we have included the numbers for the sub-giant branch (SGB). This region, not shown on Figure \ref{fig:eBSS} but with a clear branch extending from the TO point to the RGB, is included to support our idea that the RGB in these filters is not contaminated by evolved BSS. Doing the calculations to estimate the expected number of stars on the RGB we get a small difference of only 2.8\% with the number of stars observed.

\begin{table}[ht]\centering
 \caption{Times spent in different regions selected in the UV CMD (Figure \ref{fig:eBSS}) according to the MESA models, counting the stars in the complete WFC3 data set. For each evolutionary stage, the first row is the time spent in the region followed by the number of observed and expected stars. The HB for the BSS models, corresponds to the time the stars spend since the end of the green region until the end of the AGB. For the HB stars of the $0.9M\_{odot}$ model, the count includes the stars expected from the HB and AGB.} 
 \centering
 \begin{tabular}[c]{|c|cc|c|c|c|c|}
\cline{4-7}
\multicolumn{2}{c}{}& & \multicolumn{4}{c|}{{\bf Models}} \\ 
\cline{4-7}
\multicolumn{2}{c}{}& & {$0.9M_{\odot}$} & {$1.1M_{\odot}$} & {$1.4M_{\odot}$} & {$1.8M_{\odot}$}\\ 
\hline 
\multirow{3}{*}{\begin{sideways}{\bf SGB}\end{sideways}}&  
\multicolumn{2}{c|}{Time ($Myr$)} & 415 & 171 & 115 & 54 \\
\cline{2-7} &
\multicolumn{2}{c|}{Obs} & 2940 & \multicolumn{3}{c|}{N/A}\\ \cline{2-7}
\cline{2-7} &
\multicolumn{2}{c|}{Exp} & \multicolumn{4}{c|}{N/A}\\  
\hline
\hline 
\multirow{3}{*}{\begin{sideways}{\bf eBSS}\end{sideways}}&  
\multicolumn{2}{c|}{Time ($Myr$)} & N/A & 319 & 195 & 54 \\
\cline{2-7} &
\multicolumn{2}{c|}{Obs} & N/A & \multicolumn{3}{c|}{100}\\ \cline{2-7}
\cline{2-7} &
\multicolumn{2}{c|}{Exp} & \multicolumn{4}{c|}{N/A}\\  
\hline
\hline \multirow{3}{*}{\begin{sideways}{\bf RGB}\end{sideways}} &   
\multicolumn{2}{c|}{Time ($Myr$)} & 420 & 340 & 220 & 84 \\
\cline{2-7} &
\multicolumn{2}{c|}{Obs} & 3060 &  \multicolumn{3}{c|}{N/A}  \\ \cline{2-7}
\cline{2-7} &
\multicolumn{2}{c|}{Exp} & 2975 & \multicolumn{3}{c|}{$\sim 100$} \\  
\hline
\hline \multirow{3}{*}{\begin{sideways}{\bf HB}\end{sideways}} &   
\multicolumn{2}{c|}{Time ($Myr$)} & 85 & 182 & 180 & 160 \\
\cline{2-7} &
\multicolumn{2}{c|}{Obs} & 710 & \multicolumn{3}{c|}{N/A}  \\ \cline{2-7}
 & \multicolumn{2}{c|}{Exp} & 620 &  \multicolumn{3}{c|}{$\sim 100$}  \\
\hline 
 \end{tabular}
 \label{table:times-wfc3}
\end{table}

We then followed the same procedure but using the BSS models (magenta curves in Figure \ref{fig:eBSS}). The BSS population does not outline clear evolutionary stages on the CMD, besides from their position on the MS. Following the models, we identify a region between the SGB, RGB, BSS, and HB, highlighted in green, where we would expect mainly stars that have evolved from a BSS (even though these are not the only evolved BSS present on the CMD we will call this region eBSS, and for the rest of the evolutionary stages we will refer to them as the HB of the BSS, RGB of the BSS, etc.). Because of the likelihood of blends, especially in the region right next to the SGB, we have included a mild error cut in the magnitude of this sample. Considering the $1.4M_\odot$ model, we find that, for a star of this mass, the time it takes to go from its TO point to the end of the green region (before evolved BSS and normal stars share CMD space) is 200 Myr. The same time scale was found when going from the end of the green region up to the AGB phase. Considering these two time scales we would expect a contamination of 100 stars to the HB and surrounding regions from the evolution of BSS. 

The total number of expected stars from non-BSS stars in the HB region of the UV CMD is 620, in contrast, the observed number of stars is 710. The results obtained for the number of expected stars for the aforementioned region from the evolution of BSS brings the observed and expected number of stars from non-BSS stars into agreement. 

Looking now at the at the radial distributions for the four coloured regions, the BSS distribution looks similar to that which we call eBSS (green region). Table \ref{table:kstest-ebss} shows the {\it p}-values obtained from the KS-test performed between the different populations. These results support the idea that both distributions, BSS and evolved BSS, were drawn from the same population with a {\it p}-value of $0.98$. On the other hand, the HB distribution looks similar to that of the RGB but the KS-test rejects the hypothesis of these coming from the same distribution with a {\it p}-value of $\sim 10^{-3}$. Looking closely we can see that the HB star distribution appears to be slightly more centrally concentrated than the RGB stars. Doing the same for the evolved BSS we can also reject these stars coming from the same sample as the RGB or HB with {\it p}-values for the KS-test of 0.01 and 0.03.

\begin{table}[ht]\centering
 \caption[KS-test results between the populations selected on Figure \ref{fig:eBSS}]{KS-test {\it p}-value results between the populations selected in Figure \ref{fig:eBSS}}
 \centering
 \begin{tabular}{  x{40pt}  x{40pt}  x{40pt}  x{40pt}}
& & & \multicolumn{1}{c}{\cellcolor{light-gray}{\bf BSS}}            \tabularnewline \cline{4-4}
& & \multicolumn{1}{c}{\cellcolor{light-gray}{\bf eBSS}}	&	0.98		\tabularnewline \cline{3-4}
& \multicolumn{1}{c}{\cellcolor{light-gray}\bf RGB}	&	$\sim 10^{-4}$	&	$\sim 10^{-5}$	 \tabularnewline \cline{2-4}
 \multicolumn{1}{c}{\cellcolor{light-gray}\bf HB}  &  $\sim 10^{-3}$  &  0.03  & 0.01		\tabularnewline \cline{1-4}
 \end{tabular}
 \label{table:kstest-ebss}
\end{table}

\begin{figure}[H]
	\centering
	\includegraphics[width=5.in]{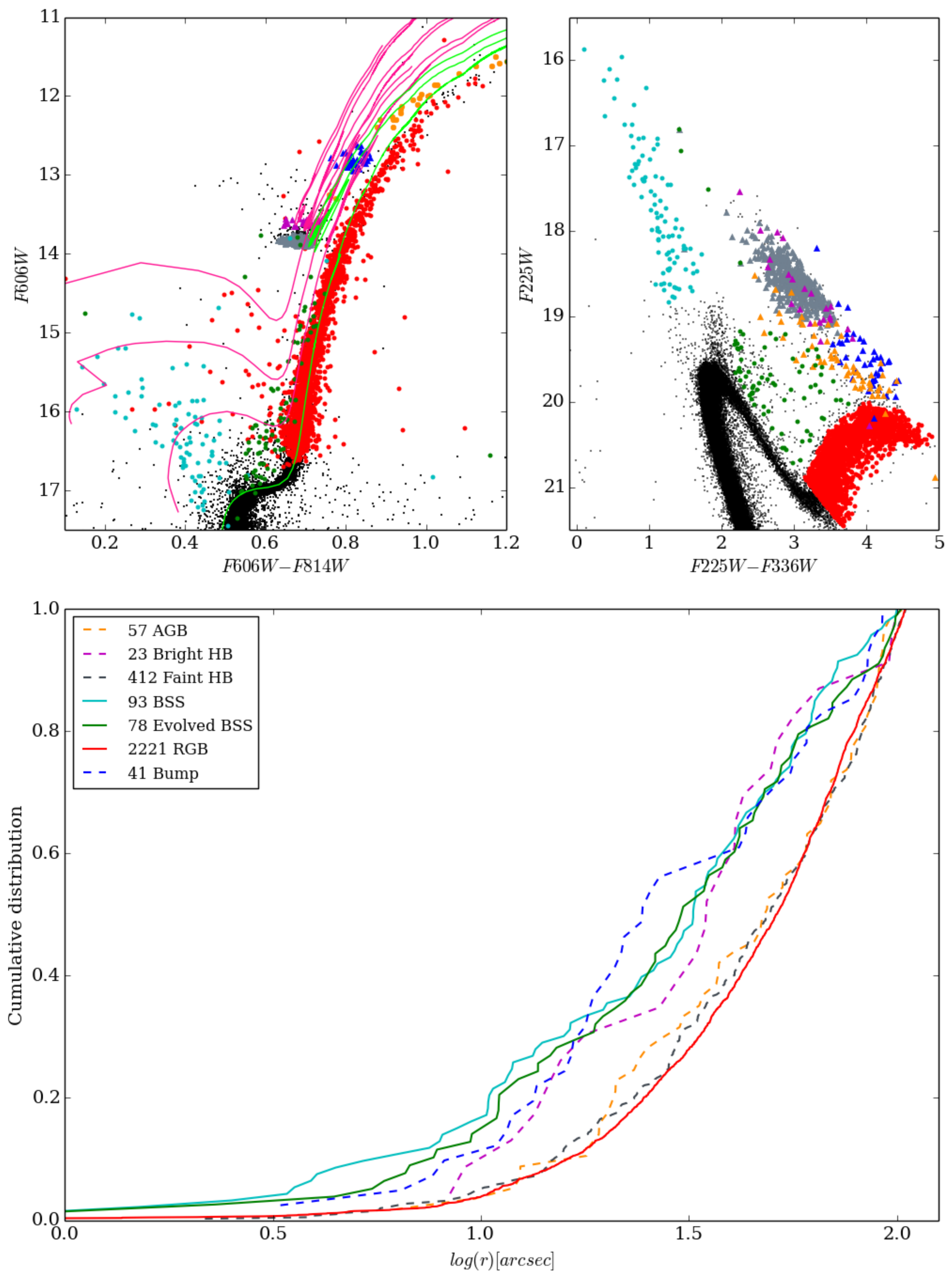}
	\caption{{\it Top-left:} $F606W,F606W-F814W$ ($V,V-I$) colour-magnitude diagram (CMD) of the core (with $r\leq105''$) of 47 Tucanae with the same MESA models as Figure \ref{fig:eBSS}. {\it Top-right:} $F225W,F225W-F336W$ ($U,U-B$) CMD for the same region. In both CMDs stars represented by triangles mean they have been selected on the $V,V-I$ CMD, coloured circles on the $U,U-B$. Different colours indicate different populations as indicated on the legend of the bottom plot. {\it Bottom:} The radial distributions of the different selected evolutionary stages.}
	\label{fig:eBSS_ATA}   
\end{figure}

To further expand the study and identification of post-MS BSS, we now compare our data to the ACS data. We can see in Figure \ref{fig:eBSS_ATA} that isolating the AGB on the UV CMD is almost impossible, but it becomes much easier on the ACS CMD, especially at the fainter end of the AGB. By cross-matching the stars from the ACS CMD to the UV, we can identify the AGB stars and obtain a cleaner sample to count the stars and to construct the radial distribution. Using this smaller data set we see that the number of stars on the HB and AGB are not consistent with the models. As shown in table \ref{table:times-acs} we expect $370$ HB stars but we observe $\sim410$, for the AGB the numbers are $80$ and $\sim100$ counting the AGB plus the bump highlighted in blue in Figure \ref{fig:eBSS_ATA}. In summary, taking into account the evolution of non-BSS related stars, we would expect 450 stars from the HB to the observable part of the AGB, not far from the observe count of 510.

\begin{table}\centering
 \caption{Time spent in different regions of the CMD according to the MESA models for the regions selected from the ACS CMD as shown in Figure \ref{fig:eBSS_ATA}. The eBSS stars were chosen from the WFC3 data. The count of stars observed on the AGB for the $0.9M_{\odot}$ model includes the stars from the bump highlighted in blue on Figure \ref{fig:eBSS_ATA}. The ages for the AGB on the BSS models ($1.1M_{\odot}$,$1.4M_{\odot}$ and $1.8M_{\odot}$) are calculated within the same magnitude range as the  $0.9M_{\odot}$ stars. For the $0.9M_{\odot}$ model the numbers for the bump region make reference to the RGB bump from the evolution of normal stars, while for the BSS models the observed number is the number of stars in the AGB bump. The number of expected stars for the $1.8M_{\odot}$ models are biased by the number of stars in the eBSS region, using the actual number of stars above the $1.4M_{\odot}$ model on the CMD only 1 star on the HB and 4 on the AGB would be expected.}
   \centering
 \begin{tabular}[c]{|c|cc|c|c|c|c|}
\cline{4-7}
\multicolumn{2}{c}{}& & \multicolumn{4}{c|}{{\bf Models}} \\ 
\cline{4-7}
\multicolumn{2}{c}{}& & {$0.85M_{\odot}$} & {$1.1M_{\odot}$} & {$1.4M_{\odot}$} & {$1.8M_{\odot}$}\\ 
\hline 
\multirow{3}{*}{\begin{sideways}{\bf eBSS}\end{sideways}}&  
\multicolumn{2}{c|}{Time ($Myr$)} & N/A & 319 & 195 & 54 \\
\cline{2-7} &
\multicolumn{1}{c}{\multirow{2}{*}{WFC3}} &
\multicolumn{1}{|c|}{Obs} & N/A & \multicolumn{3}{c|}{80}\\ \cline{3-7}
\multicolumn{1}{|c|}{}        &                &  
\multicolumn{1}{|c|}{Exp} & \multicolumn{4}{c|}{N/A}\\  
\hline
\hline \multirow{3}{*}{\begin{sideways}{\bf RGB}\end{sideways}} &   
\multicolumn{2}{c|}{Time ($Myr$)} & 420 & 340 & 220 & 84 \\ 
\cline{2-7} &
\multicolumn{2}{|c|}{Obs} & 2200 &  \multicolumn{3}{c|}{N/A}  \\ \cline{3-7}
\cline{2-7} &
\multicolumn{2}{|c|}{Exp} & N/A & 85 & 90 & 120 \\  
\hline
\hline \multirow{3}{*}{\begin{sideways}{\bf Bump}\end{sideways}} &   
\multicolumn{2}{c|}{Time ($Myr$)} & 35 &  40 & 44 & 20 \\
\cline{2-7} &
\multicolumn{1}{c}{\multirow{2}{*}{ACS}} &
\multicolumn{1}{|c|}{Obs} & 180  &  \multicolumn{3}{c|}{40}  \\ \cline{3-7}
\multicolumn{1}{|c|}{}        &                &  
\multicolumn{1}{|c|}{Exp}& 185 & 11 & 18  & 28 \\
\hline
\hline \multirow{3}{*}{\begin{sideways}{\bf HB}\end{sideways}} &   
\multicolumn{2}{c|}{Time ($Myr$)} & 70 & 68 & 60 & 51 \\
\cline{2-7} &
\multicolumn{1}{c}{\multirow{2}{*}{ACS}} &
\multicolumn{1}{|c|}{Obs} & 410 &  \multicolumn{3}{c|}{25}\\ \cline{3-7}
\multicolumn{1}{|c|}{}        &                &  
\multicolumn{1}{|c|}{Exp}& 370 & 18 & 25  & 72 \\
\hline
\hline \multirow{3}{*}{\begin{sideways}{\bf AGB}\end{sideways}} &   
\multicolumn{2}{c|}{Time ($Myr$)} & 15 & 15 & 20 & 30 \\
\cline{2-7} &
\multicolumn{1}{c}{\multirow{2}{*}{ACS}} &
\multicolumn{1}{|c|}{Obs} & 100 & \multicolumn{3}{c|}{$\sim 10$}  \\ \cline{3-7}
\multicolumn{1}{|c|}{}        &                &  
\multicolumn{1}{|c|}{Exp}& 80 &  \multicolumn{3}{c|}{$\sim 8$}  \\
\hline
 \end{tabular}
 \label{table:times-acs}
\end{table}

Going back to the evolved BSS, using the MESA models on the ACS CMD, we can now point out where the RGB and HB of the BSS fall on the CMD. The first thing we notice is that the HB for the evolved BSS is brighter than the HB for the normal stars which makes it reasonable to split this population between faint and bright HB. Another interesting result is the fact that the RGB bump for the evolved BSS falls in the same region where the AGB bump was thought to be, even with the wide range in BSS mass the RGB bump for the BSS models is always sitting just above the HB. This is the reason we have separated this group of stars from the rest of the AGB for further inspection. In this bump alone we count 41 stars. If this is indeed the RGB bump of evolved BSS we would expect to have around 20 stars coming from the evolution of BSS, which means that at least half the stars in this bump are actually evolved BSS and not AGB stars. We must mention that the numbers of expected stars for the $1.8M_{\odot}$ model obtained by starting with a star count of 80 stars on the green region is unrealistic. As we can see on the UV CMD, there are only 4 stars just above the $1.4M_{\odot}$ model. If we take this number as the actual observed count of stars we would expect a total of only 1 and 4 stars for the HB and AGB respectively.

Before splitting the HB and AGB (HB in faint and bright HB, and AGB in AGB and AGB bump) both radial distributions look more centrally concentrated than the RGB. With the samples separated as explained above, we compare the radial distributions for all the populations highlighted in Figure \ref{fig:eBSS_ATA} and report the KS-test results in table \ref{table:kstest-ebss_ata}. The bright HB, AGB bump and the BSS distributions look very similar and KS-test results show that we cannot reject the possibility that all three samples are drawn from the same population with {\it p}-values of $0.77$ for BSS against bright HB and $0.58$ for BSS versus AGB bump. Using the same statistic we find a {\it p}-value of $0.04$ and $0.08$ between the AGB and the AGB bump and for the fHB versus bHB respectively. 

In this reduced sample the relation between BSS and eBSS is similar as before, obtaining a {\it p}-value of $0.99$ when comparing their radial distributions. KS-test results also point towards the eBSS being drawn from the same populations as the bHB and AGB bump with {\it p}-values of $0.88$ and $0.64$ respectively.  


Separating the HB and AGB has also helped us to make more sense of the normal evolution of stars in the cluster. Now the cumulative radial distributions of the RGB, fHB and AGB are more closely related with {\it p}-values of $0.40$ for RGB vs. fHB, $0.89$ for RGB vs. AGB, and $0.98$ for fHB vs. AGB.

\onecolumngrid

\begin{table}[H]\centering
 \caption{KS-test {\it p}-value results between the populations selected on Figure \ref{fig:eBSS_ATA}}
 \centering
 \begin{tabular}{  x{40pt}  x{40pt}  x{40pt}  x{40pt} x{40pt}  x{40pt}  x{40pt} } 
& & & & & &   \multicolumn{1}{c}{\cellcolor{light-gray}{\bf BSS}}                    \tabularnewline \cline{7-7}

& & & & & \multicolumn{1}{c}{\cellcolor{light-gray}{\bf eBSS}}	&	0.99		     \tabularnewline \cline{6-7}

& & & & \multicolumn{1}{c}{\cellcolor{light-gray}\bf RGB} &	$\sim 10^{-4}$ &	$\sim 10^{-5}$	 \tabularnewline \cline{5-7}

& & & \multicolumn{1}{c}{\cellcolor{light-gray}\bf fHB} & 0.40 &	$\sim 10^{-3}$ &	$\sim 10^{-4}$     \tabularnewline \cline{4-7}

& & \multicolumn{1}{c}{\cellcolor{light-gray}\bf bHB} &  0.08  & 0.04 &  0.88 & 0.77  \tabularnewline \cline{3-7}

& \multicolumn{1}{c}{\cellcolor{light-gray}\bf AGB} & 0.17 & 0.98 & 0.89 & 0.06 & 0.02   \tabularnewline \cline{2-7}

 \multicolumn{1}{c}{\cellcolor{light-gray}{\bf Bump}}  & 0.04	& 0.36 & $\sim 10^{-4}$ & $\sim 10^{-3}$ & 0.64 & 0.58 \tabularnewline \cline{1-7}
 \end{tabular}
 \label{table:kstest-ebss_ata}
\end{table}

\clearpage

\twocolumngrid

\subsection{Mass Estimation}

The location of a BSS in the CMD is not necessarily representative of the mass of a single star. From the CMD and MESA models we obtained the evolutionary mass for the BSS which ranges between the TO mass and $1.8M_{\odot}$. Using the radial distance of the star from the center of the cluster we derived the dynamical mass which is the sum of the masses of all the stars in the system. The procedure to derive the dynamical masses for the BSS and BSS related populations (explained in detailed in {\bf paper1}), consists in deriving a relation between the mass (M) of a population of stars and its radial distribution (R). To achieve this, we first selected 3 samples of stars along the MS, each containing over 10000 stars. Even in this small mass range, their radial distributions show evidence of mass segregation, with the brightest and more massive sample significantly more centrally concentrated than those with lower masses (see Figure 3 in {\bf paper1}). We then calculated $R_{20}$ and $R_{50}$ for each of the samples, where $R_{20}$ and $R_{50}$ refer to the distances from the center of the cluster where the cumulative radial distributions reach 20 and 50\% respectively. Using these values together with the median mass of each MS sample (from the models), we found linear relations between $log(M)$ and $log(R)$ for both $R_{20}$ and $R_{50}$. (see equations (2) and (3) for WFC3 field, and (6) and (7) for the ACS field in {\bf paper1}). Table \ref{table:masses} shows the results for the different selected samples in both the ACS and WFC3 data. The values for the MSBn and fBSS are almost identical, the bBSS have the highest mass estimates, and we can see similar masses for the complete sample of BSS and the eBSS, bump and bHB.

The difference between the mass estimates when using the 105'' field versus the 160'' field is  expected for stellar populations in the CMD that have a wide range of mass values. Under mass segregation higher mass stars migrate towards the center of the cluster, thus, when we compare the mass estimates at $R_{20}$, the ACS field shows higher mass values because it is closer to the center of the cluster than $R_{20}$ for the WFC3 field, the same is true for $R_{50}$. The differences at $R_{20}$ and $R_{50}$ for the same field come from the same principle.

\begin{table}[ht]\centering
 \caption{Results for the mass estimation in both the WFC3 complete 160 arcseconds field and the reduced ACS field (105 arcseconds).} 
 \centering
 \begin{tabular}[c]{ |c |c |c |c |c |}
\cline{2-5}
\multicolumn{1}{c|}{}&  \multicolumn{2}{c}{$R\leq160$}& \multicolumn{2}{c|}{$R\leq105$} \\ 
\cline{2-5}
\multicolumn{1}{c|}{}&  {$M_{R20}$}& {$M_{R50}$} & {$M_{R20}$}& {$M_{R50}$}\\ 
\hline 
\multirow{2}{*}{{\bf MSBn}}&  
\multirow{2}{*}{$1.15_{-0.07}^{+0.04}$} & \multirow{2}{*}{$1.05_{-0.10}^{+0.05}$} & \multirow{2}{*}{$1.33_{-0.16}^{+0.08}$} & \multirow{2}{*}{$1.35_{-0.15}^{+0.13}$} \\ 
& & & & \\
\hline 
\multirow{2}{*}{{\bf BSS}}&  
\multirow{2}{*}{$1.47_{-0.12}^{+0.20}$} & \multirow{2}{*}{$1.22_{-0.15}^{+0.17}$} & \multirow{2}{*}{$1.95_{-0.31}^{+0.23}$} & \multirow{2}{*}{$1.44_{-0.43}^{+0.15}$} \\ 
& & & & \\
\hline 
\multirow{2}{*}{{\bf bBSS}}&  
\multirow{2}{*}{$1.64_{-0.24}^{+0.06}$} & \multirow{2}{*}{$1.43_{-0.56}^{+0.20}$} & \multirow{2}{*}{$2.38_{-0.70}^{+0.43}$} & \multirow{2}{*}{$2.27_{-1.36}^{+0.83}$}  \\ 
& & & & \\
\hline 
\multirow{2}{*}{{\bf fBSS}}&  
\multirow{2}{*}{$0.93_{-0.40}^{+0.09}$} & \multirow{2}{*}{$0.98_{-0.25}^{+0.09}$} & \multirow{2}{*}{$1.50_{-0.45}^{+0.50}$} & \multirow{2}{*}{$1.26_{-0.25}^{+0.26}$} \\ 
& & & & \\
\hline 
\multirow{2}{*}{{\bf eBSS}}&  
\multirow{2}{*}{$1.40_{-0.07}^{+0.07}$} & \multirow{2}{*}{$1.38_{-0.13}^{+0.07}$} & \multirow{2}{*}{$1.72_{-0.14}^{+0.11}$} & \multirow{2}{*}{$1.79_{-0.27}^{+0.19}$} \\ 
& & & & \\
\hline 
\multirow{2}{*}{{\bf Bump}}&  
\multirow{2}{*}{N/A} & \multirow{2}{*}{N/A} & \multirow{2}{*}{$1.52_{-0.32}^{+0.28}$} & \multirow{2}{*}{$2.10_{-0.67}^{+1.07}$} \\ 
& & & & \\
\hline 
\multirow{2}{*}{{\bf bHB}}&  
\multirow{2}{*}{N/A} & \multirow{2}{*}{N/A} & \multirow{2}{*}{$1.45_{-0.58}^{+0.27}$} & \multirow{2}{*}{$1.31_{-0.51}^{+0.26}$} \\ 
& & & & \\
\hline 
 \end{tabular}
 \label{table:masses}
\end{table}
 
\section{DISCUSSION} \label{sec:discussion}

\subsection{Blue Stragglers}
\label{subsec:dis_BSS}

The results reported in section \ref{subsec:res_BSS} indicate there are two distinct BSS sequences present within a radius of 160 arcseconds from the center of 47 Tuc. The {\it p}-value of 0.01 obtained for the KS-test between the faint and bright BSS confirms that the two populations have different distributions and therefore come from different samples, suggesting different formation mechanisms. Previous studies, including the BSS in the core of 47 Tuc, have also argued in favour of more than one formation mechanisms \citep{mapelli2004,mapelli2006,monkman2006} going on in this GC; primary stellar evolution and direct collisions. Furthermore, the estimated dynamical masses for faint and bright BSS systems are also different, with the bright BSS considerably more massive than the faint ones.

The bright BSS are very centrally concentrated (20\% of the bright BSS are within 10 arcseconds of the center, and the distribution reaches 50\% at only $\sim 30$ arcseconds), also their cumulative radial distribution does not resemble any of the other populations identifiable in the CMD. This prevents us from linking their formation to any specific group of stars. On the other hand, the mass estimate of $2.38M_{\odot}$ at $R_{20}$ using the ACS data for the bright BSS, indicates that they must come from the interactions of at least three stars, possibly through the evolution of hierarchical triple systems or encounters involving more than two stars. Even when we include the lower limit error, the mass value is still over twice the TO mass. 

In contrast, the faint BSS are less segregated towards the center but still more concentrated than most of the other populations. Their cumulative radial distribution looks very similar to that of the MSBn (\ref{fig:pop_sel}), confirmed by the 0.76 {\it p}-value obtained for the KS-test. The resemblance of their radial distributions points to a binary origin for the faint BSS. The estimated masses of these two populations are also similar, with the faint BSS being a little less massive that the MSBn as could be expected for a final product of binary evolution. 

\subsection{Evolved Blue Stragglers}
\label{subsec:dis_eBSS}

Before separating the bright and faint HB stars, the combined HB distribution looks more centrally concentrated than that of the RGB, a difference that is sustained by the KS-test results: $\sim99$\% probability that the distributions were taken from different samples. According to mass segregation, for a relaxed population of stars to be more centrally concentrated it would need to be more massive. This fact contradicts the models and theory about stellar evolution where some mass loss is expected between the RGB and HB (see \cite{origlia2007,origlia2010}, for example). Although recent results indicate that the bulk of the mass loss happens when the star is closer to the tip of the AGB (\citealt{massloss}, and references within), there is no evidence of mass gain or any other process that could lead to the HB stars being significantly more massive than the RGB stars. The models superposed on the UV CMD in Figure \ref{fig:eBSS} show that the HB and AGB of the BSS occurs to the right of the normal HB, but there are no stars in that part of our CMD. This is due to saturation of the images in the F336W filter at a magnitude of 15.25, which makes the color of any star with a magnitude above the saturation level pushed below the saturation line. In this case, the saturated stars are pushed into the same CMD position as the real HB. The contamination to the HB can also be noticed by just counting the stars in that region, which gives an observed number of stars much higher than expected. This overabundance of stars can be justified if we add the expected number of HB stars from normal evolution to the expected number of evolved BSS stars in that part of the CMD by using the time scales derived in section \ref{subsec:res_eBSS} (see table \ref{table:times-wfc3}).

The relation of the evolved BSS to the BSS was confirmed by their cumulative radial distributions with {\it p}-value of 0.98 for the KS-test. Interestingly, when we plot the stars in the green region on the ACS data we find they lie very close to, if not on top of, the SGB and RGB of the evolution of normal stars, close to the portion of the CMD that \cite{beccari2006} identified to try to find BSS starting their RGB phase. Selecting these stars on the UV CMD allows us to obtain a cleaner sample with a much lower chance of selecting normal RGB stars.

Because the number of BSS is almost the same as the number of eBSS, the time a star spends in both of these parts of the CMD also has to be similar. The time calculated for the eBSS from the models suggests a short BSS MS lifetime of $\sim 200-300$ Myr. This result disagrees with those found by \cite{sills2000} and \cite{chatt2013} (between 1 and 3 Gyr). According to \cite{lombardi1996}, collisions would produce BSS with shorter MS lifetimes, compared to stars with the same amount of hydrogen in the stellar core, in agreement with our lifetime estimations.


The same overabundance of HB stars is observed when analysing the ACS CMD. In this case when we superpose the models we noticed that the HB for the BSS is brighter than the HB for the normal evolution of stars. In fact, when we split the HB into faint and bright HB, the numbers of observed and expected stars agree. As noted by \cite{beccari2006}, we also find that the distribution of the bright HB and that of the BSS are likely drawn from the same population while the faint HB distribution resembles that of the RGB. We have used this fact to obtain a clean HB sample in the UV CMD that has allowed us to confirm the contamination of this CMD region and the predictions from our models.

From the ACS data, we can see that the overabundance of stars on the HB also extends to the AGB. Again we explained this extra population of stars by adding up the number of expected stars for the AGB and evolved BSS. According to the numbers reported we expect at least half the stars on the AGB plus AGB bump to be evolved BSS. The fact that we can statistically state that the AGB and AGB bump do not come from the same distribution but the BSS and AGB bump most likely do, supports our assumption that this bump is mostly populated by evolved BSS going up the RGB for the first time. More specifically our BSS models place the RGB bump for the BSS in the same region as the AGB bump. Having the radial distribution of the AGB without the bump agreeing with that of the RGB tells us this part of the CMD is dominated by stars coming from the evolution of normal stars. This excess of stars in the AGB was studied earlier by \cite{bailyn} and \cite{beccari2006}, who came to two different conclusions. \citeauthor{bailyn} suggested that this excess was due to BSS going through their HB stage, but according to our models and as stated by \citeauthor{beccari2006}, the HB of the BSS is much fainter than the AGB bump. \citeauthor{beccari2006} relates this contamination to the {\it ``high-mass binary by-products currently ascending the RGB for the first time"}. Our results are in good agreement with \cite{beccari2006}, but we have also been able to constrain the bulk of the contamination of the AGB bump as due to the RGB bump of the BSS.

The relation between the BSS with the eBSS, bHB and the bump, can also be seen when we compare the mass estimates of these populations.

\section{CONCLUSIONS} \label{sec:conclusion}
We have identified a large sample of over 200 BSS and evolved BSS in HST UV data of the core of 47 Tuc. Expanding our research using available data in the visible, we have studied the properties of this population including their masses, possible formation mechanisms, and their evolution.  

When we separate the bright and faint BSS we find that the bright BSS show a much more centrally concentrated radial distribution and higher mass estimates, properties that suggest an origin involving triple or multiple stellar systems. In contrast, the faint BSS are less concentrated, with a radial distribution similar to the MSBn pointing to this populations as their likely progenitors.  

Isolating a sample containing only evolved BSS had, until now, only been attempted on the HB. The evolved BSS selected on the UV CMD along with the MESA models and the agreement between the radial distributions of the BSS, evolved BSS, bright HB, and AGB bump, allowed us to construct the story of the evolution of BSS. The time scales and number of observed and expected stars agree nicely with the BSS having a post-MS evolution comparable to that of a normal star of the same mass. The disagreement between our estimated MS lifetime and those found by others indicate that a more detailed study of individual BSS properties is necessary to constrain these values. 

We have also been able to select clean samples in the different stellar evolutionary stages for the normal evolution of stars. Here we find that the cumulative radial distributions for the upper MS, RGB, faint HB and AGB, seem to all come from the same sample as expected for stars of the same mass. It is important to mention that in both the AGB and the AGB bump, we find stars from the evolution of normal stars as well as those coming from the evolution of BSS. But the number of stars and their radial distributions have allowed us to state the dominant population in each sample. 

Future studies using high quality spectra, will tell us more about the formation and evolution of BSS. Each formation mechanism leaves BSS with different chemical properties and possible companions, both of which could be identified and characterised through spectroscopy (\cite{ferraro2016} and references within).

\clearpage
\bibliography{biblio}


\end{document}